\documentclass[journal,twocolumn]{IEEEtran}

  \usepackage{graphicx}
    \usepackage{amsmath}
    \usepackage{amssymb}
    \usepackage{array}
    \usepackage{multicol}
    \usepackage{bm}
    \usepackage{multirow}

\newtheorem{lemma}{Lemma}

\begin{document}
%
\title{Stochastic Modeling and Analysis of User-Centric Network MIMO Systems}

\author{
Caiyi Zhu and Wei Yu%
\thanks{Manuscript submitted to IEEE Transactions on Communications on
February 2, 2018, revised on June 14, 2018, accepted on August 14, 2018. 
This work was supported in part by Natural Sciences and
Engineering Research Council (NSERC) and in part by Huawei Technologies Canada.
This work has been presented in part in the IEEE Workshop on Signal Processing
Advances for Wireless Communications (SPAWC), Edinburgh, U.K., July 2016.
The authors are with The Edward S.~Rogers Sr.\ Department of Electrical
and Computer Engineering, University of Toronto, Toronto, ON M5S 3G4, Canada.
Emails: caiyi.zhu@mail.utoronto.ca, weiyu@comm.utoronto.ca }
}

\markboth{}
{Zhu, Yu: Stochastic Modeling and Analysis of User-Centric Cooperation in Network MIMO Systems}

\maketitle
\begin{abstract}
This paper provides an analytical performance characterization of both
uplink (UL) and downlink (DL) user-centric network multiple-input
multiple-output (MIMO) systems, where a cooperating BS cluster is
formed for each user individually and the clusters for different users
may overlap.  In this model, cooperating BSs (each equipped with
multiple antennas) jointly perform zero-forcing beamforming to the set
of single-antenna users associated with them. As compared to a
baseline network MIMO systems with disjoint BS clusters, the effect of
user-centric clustering is that it improves signal strength in both UL
and DL, while reducing cluster-edge interference in DL. This paper
quantifies these effects by assuming that BSs and users form Poisson
point processes and by further
approximating both the signal and interference powers using Gamma
distributions of appropriate parameters. We show that BS cooperation
provides significant gain as compared to single-cell processing for
both UL and DL, but the advantage of user-centric clustering over the
baseline disjoint clustering system is significant for the DL
cluster-edge users only. Although the analytic results 
are derived with the assumption of perfect channel state information
and infinite backhaul between the cooperating BSs, they
nevertheless provide architectural insight into the design of
future cooperative cellular networks.

\end{abstract}


\section{Introduction}

Intercell interference is the main limiting factor in the
physical-layer of modern wireless cellular networks with densely
deployed base-stations (BSs).  Network multiple-input multiple-output
(MIMO) is a promising technique for interference mitigation in which
BSs jointly transmit information to and receive information from the
multiple users via coherent beamforming across multiple BSs
\cite{NetworkMIMO_review}. This paper aims to provide analytic
modeling and performance characterization of such network MIMO
systems.

Although the elimination of all intercell interference would in
theory require cooperating BSs across the entire network, such
complete cooperation is clearly impractical due to the computational
complexity, delay and the capacity constraints of the backhaul, and
likely not necessary as distant BSs make little impact on user's
signal strength or interference power.
Thus, practical implementations of network MIMO systems would involve
only limited set of BSs forming cooperation clusters of finite sizes.
A natural way to limit the cooperation size is to partition all the BSs in
the network into disjoint clusters, so that BSs in each cluster
jointly serve all users associated with all the BSs in the cluster.
This is akin to a distributed antennas system, but in such an
architecture significant inter-cluster interference still exists for
cluster-edge users.

Instead of the baseline network MIMO system above, this paper focuses
on a \emph{user-centric clustering} strategy where a BS cooperation
cluster is formed for each user individually, and clusters for
different users can partially overlap.  Such a clustering strategy
allows the user to be always placed at the center of its cluster, so
that signal strength is improved and inter-cluster interference
reduced.

Although the performance advantage of user-centric clustering is
intuitive, the mathematical analysis of such an architecture is a
challenging task. The main goal of this paper is to provide a
performance analysis of the user-centric network MIMO architecture and
to compare it quantitatively with the baseline disjoint clustering in both uplink (UL)
and downlink (DL).  Toward this end, we assume the use of
zero-forcing beamforming (ZFBF) strategy with equal power allocation
across the beams, and adopt a stochastic geometry model of the BS and
user locations and a Gamma distribution approximation of the direct
and interfering channel strength in order to facilitate an analytic
characterization of the average user rate in a user-centric network
MIMO systems as a function of network parameters such as the
cooperation cluster size.

Throughout the paper, we assume that perfect and instantaneous channel
state information (CSI) is available at the BSs in order to facilitate
analysis.  The aim of this paper is to quantify the benefit of BS
cooperation before CSI acquisition cost is taken into account.



\subsection{Related Work}

Network MIMO has long been advocated as being capable of treating inter-cell
interference as useful signals, thereby significantly improving the throughput
of wireless cellular networks \cite{somekh2007sum, jing2008multicell}. But the
existing performance evaluation of network MIMO systems has been mostly carried
out either using simplified Wyner model or by simulation; and most earlier
works have focused on the optimization of transmit strategies for network MIMO systems.
For example, for network MIMO systems with disjoint clustering,
earlier work
\cite{huang2009increasing} numerically studies the throughput performance;
\cite{zhang2009networked, jiang2012capacity} present optimization strategies
where users close to cluster edge are served with inter-cluster coordination
and users close to cluster center are served only with intra-cluster BS
coordination.
Likewise for user-centric clustering, first proposed in
\cite{papadogiannis2008dynamic}, most existing works are based on numerical
investigation and optimization strategies that select the best serving cluster
of BSs for each user \cite{ng2010linear, Binbin_SparseBF,suuser}.

The performance analysis of network MIMO systems is a challenging task, because the cooperating
BSs in a network MIMO system typically have different path-loss to the user, so
traditional analytic tools for MIMO system, such as the random matrix theory
\cite{tulino2004random}, are not ideally suited for analyzing the network MIMO
system performance, unless certain symmetry and simplifying assumptions are
adopted \cite{huh2012network}.



To account for the distance dependent path-loss in wireless communication
networks, stochastic geometry has emerged as a powerful tool for
analyzing wireless networks with \emph{random deployment} of BSs and users
that are assumed to form Poisson point processes.
Stochastic geometry is first applied to the analysis of cellular networks
with single-cell processing in term of coverage probability for both the DL
\cite{JeffAndrews_A_tractable_approach} and the UL \cite{Analytical_modeling_UL}.

The stochastic geometry analysis of coordinated multicell network is more
challenging. Toward this end, most existing analyses consider the so-called
\emph{noncoherent} joint transmission scheme \cite{tanbourgi2014tractable,
garcia2014coordinated}, where multiple BSs jointly transmit data to the same
user. Moreover, \cite{COMP_HetNets} studies the power gain in the form of coverage
probability for both coherent and noncoherent joint transmission;
\cite{sakr2014location} investigates the outage probability of both coordinated
multi-point (CoMP) non-CoMP schemes in a 2 tier network. These papers reveal
power gain from joint transmission, but
the overall system does not benefit from multiuser spatial multiplexing gain. 
Stochastic geometry can be readily applied to the analysis of outage and
coverage probabilities of these noncoherent systems, because the signal power
at a user is simply the sum of powers from the cooperating BSs without the
need to account for the effect of multicell beamforming.

This paper aims to provide a stochastic geometry analysis of \emph{coherent}
joint transmission scheme in a network MIMO system, where significant spatial
multiplexing gain can be realized via multicell beamforming. The performance
analysis for such a system is considerably more complicated, because of the need
to model the effect of beamforming. Toward this end, \cite{heath2011multiuser,
seifi2014joint} propose a series of techniques that allow an approximate
characterization of the effect of zero-forcing beamforming for multicell networks.
This enables a subsequent stochastic geometry analysis of
network MIMO systems to be carried out \cite{Kianoush_Cooperation_gain}.
In particular, instead of a typical coverage probability analysis,
\cite{Kianoush_Cooperation_gain} provides a derivation of a more useful ergodic
sum rate expression for a DL network MIMO system with disjoint clustering.


This paper further extends the analysis of network MIMO system in
\cite{Kianoush_Cooperation_gain} to the user-centric clustering case and to UL
scenarios. The user-centric case is
more complicated, because the beamforming vectors overlap with each other.
The analytic performance characterizations of this paper help illustrate the
benefit of user-centric clustering as compared to disjoint clustering for both
UL and DL scenarios.



It should be emphasized that the user-centric scheme considered in this paper
is not the only way to improve the baseline disjoint scheme with respect to the
cluster-edge user performance. Intelligent scheduling algorithms can be used to
partition the time, frequency and spatial resources together with interlaced
clustering to achieve a similar benefit \cite{caire_downlink, huh2011achieving,
ratnam_molisch_caire}.  For simplicity, this paper considers only the baseline disjoint
clustering as the reference system for comparison. The performance analysis
techniques developed in the paper would be useful for the analysis of systems
developed in \cite{caire_downlink, huh2011achieving, ratnam_molisch_caire} as well.

The network MIMO system considered in this paper is different from yet
another type of multicell cooperation scheme known as interference
nulling. The interference nulling scheme utilizes extra spatial
dimensions at the BS to create spatial nulls at selective out-of-cell
user locations, thereby eliminating selective dominant interference.
This can be done without the exchange of user data between the BSs,
only the CSI---but at the expense of additional BS antennas.
Stochastic analysis of interference nulling scheme is somewhere easier
than that of network MIMO, because the beamformed signals always come
from the same BS, rather than from multiple BSs as in network MIMO.
Works in this area include the analysis of outage probability
\cite{huang2013analytical}, optimization of interference nulling range
\cite{li2015user}, and coverage probability for two-tier networks
\cite{wu2015user}.

As mentioned earlier, the proposed analysis does not include the effect of CSI
estimation error and finite backhaul capacity. We refer readers to, e.g.,
\cite{caire2010rethinking, zhang2013downlink, lozano2013fundamental},
for a discussion on the performance of network MIMO system
considering the overhead of CSI acquisition and the effect of backhaul limitation.

\subsection{Main Contributions}

The main contributions of this paper are as follows:

\begin{enumerate}

\item This paper provides computationally efficient user rate expressions for both
UL and DL network MIMO systems under either disjoint clustering or user-centric
clustering. We adopt a Poisson model for BS and user locations, and utilize
techniques developed in \cite{Kianoush_Cooperation_gain}
for approximating the zero-forcing (ZF) signal and interference strength.

\item This paper analytically quantifies the gain of BS cooperation as a
function of the cooperation cluster size. We show that the network MIMO gain (with
either disjoint or user-centric clustering) is typically larger in the
DL than in the UL, due to the fact that the BSs have a much larger
transmit power budget than the handsets, so DL transmission is more
interference limited than UL transmission.

\item This paper shows that in term of average user rate across the
network, the relative advantage of user-centric clustering over the
baseline na\"{i}ve disjoint clustering strategy is about 15\%-20\% in the UL, and
smaller in the DL. In the UL, user-centric clustering essentially places every
user in the center of the cluster, thus resulting in both stronger signal power
and less inter-cluster interference for all UL users.

\item In the DL, user-centric clustering also brings stronger signal power
to all users, but its effect on inter-cluster interference depends on user
location. By removing the cluster edge, user-centric clustering reduces
inter-cluster interference, hence bringing in higher throughput for the
low-rate users.
(For the high-rate users, user-centric clustering actually increases
interference by introducing extra intra-cluster interference, 
but this effect is small.) The main benefit of user-centric architecture in the
DL is therefore the significant improvement in low-rate user performance.

\item Finally, this paper shows that even in the limit of unrealistically large
cluster sizes, the average user rate of a network MIMO system does not approach
that of an isolated interference-free single cell. There is significant
penalty due to ZFBF in cooperating across multiple BSs 
regardless of the cooperation cluster size.

\end{enumerate}

\subsection{Structure of the Paper and Notations}

The remainder of the paper is organized as follows. Section II describes the system model of this paper, more specifically, it provides the setting of user-centric clustering and disjoint clustering, the ZFBF design in both schemes, and the definitions of intra-cluster and inter-cluster users.  Section III gives intuitive explanation on the gain of user-centric clustering over disjoint clustering. Section IV gives the stochastic model. Section V provides the analytical result of ergodic user rate of user-centric clustering for both UL and DL using tools from stochastic geometry, which is the main mathematic contributions of this paper. Section VI provides similar analysis for disjoint clustering. Section VII makes comparison of signal and interference power between user-centric and disjoint clustering for both UL and DL based on intermediate analytical result from the previous section.  Section VIII shows numeric comparison between user-centric clustering and baseline disjoint clustering. Section IX concludes the paper.

This paper uses lower-case letter $h$ to denote scalers, bold-face lower-case
letter $\bm{h}$ for vectors, and bold-face upper-case letter $\bm{H}$ for
matrices. We use $\bm{I}_n$ to denote an $n \times n$ identity matrix, and use
$\bm{H}^H$ to denote Hermitian transpose and
$\bm{H}^{\dagger}=(\bm{H}^H\bm{H})^{-1}\bm{H}^H$ to denote the left
pseudo-inverse of a matrix $\bm{H}$. Upper-case letters such as $K$, $M$ represent
scalar constants. The expectation of a random variable is denoted as
$\mathbb{E}[\cdot]$; the $l_2$ norm of a vector is denoted as $\| \cdot\|_2$.

\section{System Model}


Consider a wireless cellular network in which each BS is equipped with $M$
antennas and each user is equipped with a single antenna.  The users are
associated with the closest BS by distance. We assume that the user density is much larger than the BS density, so that each BS has many associated users. Among all its
associated users, each BS schedules $K < M$ users in each time slot.
The ratio $K/M$ is called the loading factor.  We assume
round-robin scheduling for simplicity. We assume flat-fading channels with
full frequency reuse, i.e., transmissions from neighboring cells cause mutual
interference to each other.

\begin{figure}
\centering
\includegraphics[width=0.5\textwidth, trim=30 30 30 30, clip]{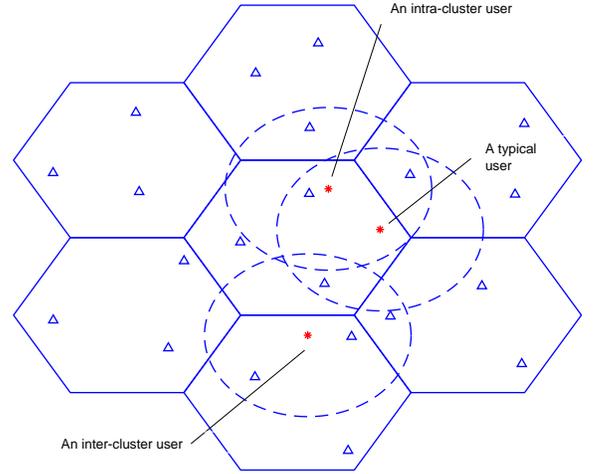}
\caption{Disjoint vs. User-centric clustering: The BSs are denoted by
triangles, the users by stars. Under disjoint clustering, the serving BSs
form non-overlapping cooperative clusters, shown in this example by the
hexagonal region. Under user-centric clustering, the serving BSs are formed
for each user individually, shown in this example by the dotted circles.}
\label{system_model}
\end{figure}

This paper aims to analyze the performance of network MIMO systems in which
each user is jointly served by a cooperative cluster of BSs.  In the
\emph{disjoint clustering} scheme, the set of BSs are partitioned into disjoint
clusters; each user is served by the cluster of BSs to which its associated BS
belongs.  In \emph{user-centric clustering}, each scheduled user forms an
individually chosen cluster of serving BSs; the
clusters for different users can partially overlap. Fig.~\ref{system_model}
illustrates the disjoint and user-centric clustering topologies.

Let $\Theta_i$ denote the cooperative cluster of serving BSs for user $i$.
Let $B_i$ be the cluster size, i.e., $B_i = |\Theta_i|$.  Recall that each
BS in $\Theta_i$ schedules $K$ of its associated users.  We denote the set
of all users scheduled by the BSs in user $i$'s serving cluster as
$\Omega_i$, so that $|\Omega_i| = K B_i$.  In the rest of the paper,
user $i$ is called the typical user.  The other $KB_i-1$ users in
$\Omega_i$ are called intra-cluster users of user $i$. All the rest of the
scheduled users in the network are termed inter-cluster users.

This paper assumes the use of ZFBF in both UL and DL, in which a beamformer for
the typical user $i$ is designed across its serving BSs in $\Theta_i$ to
null interference from/to all intra-cluster users. ZFBF is easy to implement; it
performs well at high SNR. Further, the ZF assumption significantly simplifies
the analysis, because the signal and the interference can be analyzed
separately under ZFBF.

Below, we describe the ZFBF design in both UL and DL.
The same design procedure applies to both disjoint and user-centric
clustering---the only difference being the choice of cooperative BS cluster
$\Theta_i$ for each user $i$. Throughout this paper, perfect channel state
information (CSI) is assumed to be available for the purpose of beamformer
design.

\subsection{Uplink ZFBF}

In uplink ZFBF, the message from the typical user $i$ is jointly decoded across
the BSs in $\Theta_i$, while interference from all intra-cluster users in
$\Omega_i$ is nulled. Let $\bm{y}_i \in \mathbb{C}^{MB_i}$ be the
received signal across all the serving BSs of user $i$:
\begin{eqnarray}
\bm{y}_i & = & \sum_{j}^{}\bm{h}_{ij}x_{j} + \bm{z} \\
& = & \underbrace{\bm{h}_{ii}x_{i}}_{\rm signal} +
\underbrace{\sum_{m:\; m\ne i,\; m\in \Omega_i}\bm{h}_{im}x_{m}}_{
\tiny \begin{array}{c}{\rm intra-cluster}\\ {\rm interference}\end{array}}
+
\underbrace{\sum_{j\notin \Omega_i}\bm{h}_{ij}x_{j}}_{
\tiny \begin{array}{c}{\rm inter-cluster}\\ { \rm interference }\end{array}}
+
\bm{z}  \nonumber
\label{received_sig_UL}
\end{eqnarray}
where $\bm{h}_{ij}^H=[\cdots \bm{g}_{bj}^H \cdots ]_{b\in{\Theta_i}}$
denotes the collective vector channel between user $j$ and the set of serving
BSs of user $i$, and $\bm{g}_{bj} \in \mathbb{C}^{M}$ denotes the channel
between user $j$ and BS $b$.  Here, $x_{j}$ is the transmit signal of user $j$
with power normalized to 1, i.e., $\mathbb{E}[x_{j}^2]=1$. Finally, $\bm{z}
\sim \mathcal{CN}(0, \sigma_u^{2} \bm{I}_{MB_i})$ is the background noise at
the BSs including thermal noise and other possible sources of interference and
scaled to account for transmit power normalization.

The ZF receive beamformer for user $i$ is designed to be orthogonal
to the transmission from the $KB_i-1$ intra-cluster users, so that the
interference from these users is completely eliminated.
In particular, the normalized ZF receive beamformer for user $i$ is chosen to
be the following: 
\begin{equation}\label{ZFBF_UL}
\bm{w}_{i}=\frac{(\bm{I}_{MB_i}-\bm{H}_{-i}\bm{H}_{-i}^{\dagger})\bm{h}_{ii}}
{\|(\bm{I}_{MB_i}-\bm{H}_{-i}\bm{H}_{-i}^{\dagger})\bm{h}_{ii}\|_2},
\end{equation}
where $\bm{H}_{-i} = [ \cdots \bm{h}_{ij} \cdots]_{j\neq i,j \in \Omega_i}$
denotes the channel matrix between the serving BSs of user $i$ and its
$KB_i-1$ intra-cluster users. It is easy to see that the row space of the
matrix $(\bm{I}_{MB_i}-\bm{H}_{-i}\bm{H}_{-i}^{\dagger})$ is the left null space
of $\bm{H}_{-i}$. By projecting the direct channel $\bm{h}_{ii}$ onto the
left null space of $\bm{H}_{-i}$, the signal power is maximized while the required
orthogonality is guaranteed.


The signal-to-interference-and-noise ratio (SINR) of user $i$ can now be
stated as follows:
\begin{equation}\label{SINR_UL}
\gamma_{i}=\frac{|\bm{w}_{i}^H\bm{h}_{ii}|^2}{\sum_{j \notin \Omega_i}|\bm{w}_{i}^H\bm{h}_{ij}|^{2}+\sigma_u^{2}}.
\end{equation}

\subsection{Downlink ZFBF}


In downlink ZFBF, the message for the typical user $i$ is jointly encoded across
the BSs in $\Theta_i$ using a beamforming vector $\bm{w}_i \in \mathbb{C}^{MB_i \times 1}$,
while ensuring the nulling of interference at all intra-cluster
users in $\Omega_i$. The received signal at the user $i$ is
\begin{eqnarray}
y_{i} & = & \sum_{j}^{}\bm{h}_{ji}^H\bm{w}_{j}x_{j} + z \nonumber \\
& = & \underbrace{\bm{h}_{ii}^H\bm{w}_{i}x_{i}}_{\rm signal}+
\underbrace{\sum_{m:\; m\neq i,\; i\in \Omega_m}\bm{h}_{mi}^H\bm{w}_{m}x_{m}}_{\tiny \begin{array}{c} {\rm intra-cluster} \\
{\rm interference} \end{array}} \nonumber \\
	& & \qquad +
\underbrace{\sum_{j:\; i\notin \Omega_j}\bm{h}_{ji}^H\bm{w}_{j}x_{j}}_{\tiny \begin{array}{c} {\rm inter-cluster} \\
{\rm interference} \end{array}} + z 
\label{received_sig_DL}
\end{eqnarray}
where the channel vectors are as defined in uplink,
$x_{j}$ is the signal intended for user $j$ with transmit power normalized as
$\mathbb{E}[x_{j}^2]=1$, and $z \sim \mathcal{CN}(0,\sigma_d^2)$ is background
noise with power normalization accounted for.  Note that due to the different
transmit powers, the normalized background noise powers are different in UL and DL.

The ZF transmit beamformer $\bm{w}_i$ in the downlink is exactly the same as in
uplink, i.e.,
\begin{equation}
\bm{w}_{i}=\frac{(\bm{I}_{MB_i}-\bm{H}_{-i}\bm{H}_{-i}^{\dagger})\bm{h}_{ii}}
{\|(\bm{I}_{MB_i}-\bm{H}_{-i}\bm{H}_{-i}^{\dagger})\bm{h}_{ii}\|_2}.
\end{equation}
As in the UL, the above choice of DL transmit ZF beamformer is designed to be
orthogonal to the channels to all the intra-cluster users, while maximizing the
projected signal power.
The DL SINR of user $i$ can likewise be written down as follows:
\begin{equation}\label{SINR_DL}
\gamma_{i}=\frac{|\bm{h}_{ii}^H\bm{w}_{i}|^2}{\sum_{j:\; i\notin \Omega_j}|\bm{h}_{ji}^H\bm{w}_{j}|^{2}+\sigma_d^{2}}.
\end{equation}



\section{User-Centric vs. Disjoint Clustering}

The main objective of this paper is to provide a performance analysis of the
user-centric network MIMO scheme in order to quantify its advantage as compared
to baseline disjoint clustering. Toward this end, we first give some intuitive
comparison of the signal and interference powers under the different clustering
strategies, then provide a statistical model of network MIMO system in the next
section.

\subsection{Signal Power}

In user-centric clustering, each user is always at the center of its serving
BSs. Since the channel strength is a function of the distance, the signal power
in the user-centric case is equivalent to that of a cluster-center user in
disjoint clustering, and much larger than that of a cluster-edge user. Thus on
average, user-centric clustering has an advantage in term of signal power as
compared to disjoint clustering. This holds for both UL and DL.

\subsection{UL Interference Power}

In the UL, ZFBF receiver is designed to null interference from all intra-cluster users 
for both user-centric and disjoint clustering;
in fact, the received vector of interference at the serving BSs in the two cases are
the same. However, the interference powers in the user-centric and the disjoint
clustering cases are not identical, because the interference power
is the inner product of the interference vector and the receive beamformer, and
the beamformers for the user-centric and disjoint cases are different.
Nevertheless, this effect is relatively minor.
Overall, since UL user-centric clustering gives stronger signal power and
about the same amount of interference for all users, we expect uniform
improvement in UL user rate as compared to disjoint clustering.

\subsection{DL Interference Power}

In the DL, the transmit ZF beamformer for the typical user is designed to be
orthogonal to the channels to all its intra-cluster users. Thus, the
transmission to the typical user only interferes with its inter-cluster users.
Conversely, the interference at the typical user comes only from transmission
to the users for whom the typical user is an inter-cluster user.  This observation gives distinct interference characteristics for
disjoint and user-centric clustering.

Under baseline disjoint clustering, since all users within the cluster are served by
the same set of BSs, the interference at any of the users comes only from
BSs outside of the cluster. In particular, cluster-edge users see strong
interference, while cluster-center users see less interference.

Under user-centric clustering, the serving BSs of the typical user can
overlap with the serving BSs of the interfering users.
Thus, the interference at the typical
user can come from not only BSs outside of its serving cluster, but also
from within the user's own serving cluster of BSs. As a consequence, a typical user in user-centric clustering sees more
interference than the cluster-center users in disjoint clustering; but it
sees less interference than cluster-edge users in disjoint clustering. Putting
it another way, as compared to baseline disjoint clustering, user-centric clustering
reduces interference for cluster-edge users at the expense of cluster-center users.

To summarize, DL user-centric clustering improves the cluster-edge user rate in
baseline disjoint clustering, because it both improves the signal power and reduces
interference. However, the effect of user-centric clustering for cluster-center
users is not so clear cut, as both signal and interference powers are
increased.  

\section{Stochastic Network Model}

In order to provide an analytic performance characterization of network MIMO
systems, this paper proposes a statistical model of cellular networks
accounting for both the random geographic locations of the BSs and the users,
as well as the random fluctuation of the channels, i.e., fading.

The channel from BS $b$ to user $i$ is modeled as $\bm{g}_{bi} =
\sqrt{\beta_{bi}} \bm{f}_{bi} \in \mathbb{C}^{M\times1}$. The
distance-dependent path-loss component is modeled as
$\beta_{bi}=(1+r_{bi}/d_0)^{-\alpha}$, where $r_{bi}$ is the distance between
the BS $b$ and user $i$, $d_0$ is a reference distance, and $\alpha$ is the
path-loss exponent. The Rayleigh fading component is modeled as $\bm{f}_{bi}
\sim \mathcal{CN}(0,\bm{I}_M)$.

We use tools from stochastic geometry to account for the path-loss and use a
Gamma distribution approximation to analyze the overall performance. To
facilitate the analysis, we make the following further assumptions:

\begin{enumerate}

\item The BSs are randomly placed over a two-dimensional (2D) plane as a homogenous
Poisson Point Process (PPP) with a fixed intensity $\lambda_b$, denoted as
$\Phi_b$. The PPP model allows us to derive expressions of system
performance by averaging over the random BS locations using tools from
stochastic geometry. The PPP model also accounts for the randomness of BS
deployment in practice. The validity of the PPP model for performance analysis
of cellular networks have been discussed in the literature (see e.g., \cite{JeffAndrews_A_tractable_approach}).

\item The users are also randomly placed in the 2D plane as a PPP. In our model,
the users are associated with the closest BS; each BS schedules $K$ active
users from the set of associated users. Technically the active users no longer
form a PPP. But to enable the averaging over user locations, we further
approximate that the active users form a PPP, $\Phi_u$,
with intensity $\lambda_u = K \lambda_b$.

\item In user-centric clustering, each user chooses its BS cooperation cluster
based on the distance between the BSs and the users. In particular, the user $i$'s
BS cooperation cluster $\Theta_i = \Phi_b \cap \mathcal{B}_{x_u}(R)$,
where $\mathcal{B}_{x_u}(R)$ denotes a unit circle of radius $R$ centered at user $i$
whose location is denoted as $x_u$.

\item In baseline disjoint clustering, we partition the 2D plane using a hexagonal
lattice; the BSs located within each hexagon form a cluster. In the analysis of
disjoint clustering, the hexagon shape is further approximated by a circular
disk of equal area.

\item ZFBF is used in both uplink and downlink to serve the target user, while
cancelling interference from and to the other $K B_i-1$ users scheduled by the $B_i$
BSs in the cooperating cluster. Nulling the interference for these intra-cluster users is intuitively a good design since they are in close proximity to the typical user.
We do not perform power control in either uplink or downlink. In the uplink, all users transmit at a fixed power; in the
downlink, each downlink beam transmits at a fixed power. This power
allocation model does not guarantee per-BS power constraint, but
accounting for per-BS power would make analysis much more complicated.

\item In computing interference for the typical user at the origin in the user-centric clustering case as well as for any user in disjoint clustering, we further assume that the interfering users are simply all users located outside of a circular area of radius $R$.
This is an approximation, e.g., in the downlink it assumes that the users for whom the target user is its inter-cluster user are exactly ones further than distance $R$. The approximation is accurate when the cluster size is much larger than the coverage size of each BS.
With this approximation, in the uplink, the interfering users are located outside of
$ \Phi_u \cap \mathcal{B}_o\left(R\right)$.
In the downlink, the interfering signals are the beamformed signals from the serving cluster of BSs for users outside of
$ \Phi_u \cap \mathcal{B}_o\left(R\right)$.
Note that the serving cluster can overlap with the serving BSs of the typical user.
\end{enumerate}

Note that in the BS clustering scheme described above, the number of BSs in the
cooperating cluster is not a constant, but a Poisson random variable. The later
simulation part of the paper examines the effect of having random rather than
fixed number of cooperating BSs in the cluster.

These modeling assumptions are made for the purpose of deriving efficiently computable
achievable rate expressions for network MIMO systems.  These assumptions have been used in \cite{Kianoush_Cooperation_gain} for deriving
the ergodic rate of the DL network MIMO system with disjoint clustering. This
paper extends the analysis to the UL and in particular to the user-centric
cases.





\section{Stochastic Analysis of Ergodic User Rate for User-Centric Network MIMO}

In a stochastic model of the user-centric network MIMO system with
uniformly random deployment of BSs and users, the users are all essentially
statistically identical.
In this section, we focus on a typical user
indexed as user $1$, centered at the origin. We first derive distributions of
signal and interference powers for a given realization of BS and user locations.
We then obtain an ergodic rate expression by averaging over the BS and user
location PPPs. The techniques used here are due to
\cite{Kianoush_Cooperation_gain}, where the DL disjoint case
is analyzed. The analysis here accounts for the different channel topology
and interference characteristics of user-centric cooperation.



\subsection{Analysis of Uplink}

\subsubsection{Signal Strength}
The channel strength of the intended signal for the typical user is:
\begin{equation}\label{equ_chn_sig_UL}
\begin{split}
\| \bm{h}_{11}\|^2 & 
= \sum_{b\in \Theta_1}\bm{g}_{b1}^{H}\bm{g}_{b1}\\
\qquad &=  \sum_{b\in \Theta_1}\beta_{b1}\bm{f}_{b1}^{H}\bm{f}_{b1}\sim
\sum_{b\in \Theta_1}\Gamma\left(M,\beta_{b1}\right).
\end{split}
\end{equation}
Since the entries of the MIMO channels are Gaussian distributed, the overall
magnitude of the channel between the typical user and its set of cooperating
BSs is a sum of Gamma random variables with different scale parameters
depending on the distances between the user and the BSs. We proceed to
approximate the above distribution into a form amendable to stochastic geometry
analysis. The series of approximations below are developed in part in the
analysis in \cite{heath2011multiuser, seifi2014joint,
Kianoush_Cooperation_gain}.

The analysis first uses a technique pioneered in \cite{heath2011multiuser} for
approximating the sum of Gamma distributions as a single Gamma distribution
with shape and scale parameters determined by matching the first and second order moments.

\begin{lemma}
Let $Y_i \sim \Gamma\left(k_i,\theta_i\right)$, $i=1 \cdots n$ be independent
Gamma distributed with different shape and scale parameters $k_i$ and
$\theta_i$, respectively. Consider $Y=\sum_i Y_i$, then $Y$ has the same first
and second moments as a Gamma random variable $\Gamma(k,\theta)$ with shape and
scale parameters
\begin{equation}
k=\frac{\left(\sum_{i=1}^{n}k_i\theta_i\right)^2}{\sum_{i=1}^{n}k_i\theta_i^2},
\qquad
\theta=\frac{\sum_{i=1}^{n}k_i\theta_i^2}{\sum_{i=1}^{n}k_i\theta_i}.
\end{equation}
\end{lemma}

For the channel $\| \bm{h}_{11} \|^2$ in (\ref{equ_chn_sig_UL}), this
approximation leads to $\|\bm{h}_{11}\|^2 \sim \Gamma(k_1,\theta_1)$ with
\begin{equation}
k_{1}=M\frac{\left(\sum_{b\in \Theta_1}\beta_{b1}\right)^2}{\sum_{b\in
\Theta_1}\beta_{b1}^2}, \quad
\theta_{1}=\frac{\sum_{b\in \Theta_1}\beta_{b1}^2}{ \sum_{b\in \Theta_1}\beta_{b1}}.
\label{chn_sig_power_UL}
\end{equation}

To obtain signal power, we need to further project the channel vector onto
the beamforming vector. The exact signal power distribution resulting from such a
projection is not easy to characterize. Instead, we adopt a second approximation
by drawing a parallel with the following fact on the projection of an isotropic
channel vector to a lower dimensional space (although our actual channel is not
isotropic).


If a channel vector $\bm{h} \in \mathbb{C}^N$ were isotropic in the $N$-dimensional
space such that $\|\bm{h}\|^2\sim \Gamma(N,\theta)$, then the projection of
$\bm{h}$ onto a $P$-dimensional subspace results in a Gamma distribution
$\Gamma(P,\theta)$. In other words, the shape parameter is scaled by $P/N$,
while the scale parameter is kept the same.

To obtain an approximate signal power distribution after the projection,
we apply the same scaling of the shape parameter even when the channel
is non-isotropic.  This same approximation technique is also used in
\cite{seifi2014joint, Kianoush_Cooperation_gain}.
Specifically in our case, $\|\bm{h}_{11}\|^2\sim \Gamma(k_1,\theta_1)$.  To
project the channel vector onto the ZF beamforming vector, we note that the
receive beam of the user lies in the null space of the subspace spanned by the
$K B_1-1$ interfering channel vectors.  Therefore, the shape parameter for the
signal power after projection must be scaled by $\frac{MB_1-KB_1+1}{MB_1}$.
The signal power distribution can therefore be approximated as:
\begin{equation}
\begin{split}
&\zeta_{1}^{(UL)}=|\bm{w}_{1}^H\bm{h}_{11}|^{2}\sim \Gamma\left(\frac{MB_1-KB_1+1}{MB_1}k_{1},\theta_{1}\right).\\
\end{split}
\label{sig_dist_UL}
\end{equation}
Recall that the number of BSs in the cluster $B_1$ is a Poisson random variable
with mean $\bar{B}=\lambda_b\pi R^2$. To make the analysis tractable, we
replace $B_1$ by its mean $\bar{B}$ as a further approximation.

Finally, the distribution above has parameters that depend on the BS location.
To facilitate a stochastic geometry analysis, we decompose the above signal
distribution as a linear combination of independent Gamma distributions. Using
again the technique of matching the first and second moments matching, the signal
power can now be approximated as follows \cite{Kianoush_Cooperation_gain}:
\begin{equation}
\zeta_{1}^{(UL)}=|\bm{w}_{1}^H\bm{h}_{11}|^{2}\approx\sum_{b\in
\Theta_1}\beta_{b1} G_{b1}^{(\varpi)}
\label{sig_dist_UL2}
\end{equation}
where $G^{(\varpi)}_{bj}$ are i.i.d. random variables distributed as
$\Gamma(\varpi,1)$, where $\varpi = \frac{M\bar{B}-K\bar{B}+1}{\bar{B}}$.
Here, we also use the fact that if $X\sim \Gamma\left(k,\theta\right)$, then
$cX\sim\Gamma\left(k,c\theta\right)$ for any positive $c$.

\subsubsection{Interference Strength}

As intra-cluster interference is eliminated with ZF receiver, the residual
interference only comes from inter-cluster users. In deriving the distribution
of aggregate interference, we first investigate the interference from a single
user $j$, then sum up the interference over all interfering users.

Similar to the analysis of (\ref{chn_sig_power_UL}), the interfering channel
strength can also be approximated as a Gamma random variable using the moment
matching technique as follows:
\begin{equation}
\|\bm{h}_{1j}\|^2=\sum_{b\in\Theta_1}\bm{g}_{bj}^H\bm{g}_{bj}\sim\sum_{b\in \Theta_1}\Gamma\left(M,\beta_{bj}\right)\approx \Gamma\left(k_{1j},\theta_{1j}\right),
\end{equation}
where
\begin{equation}
k_{1j}=M\frac{\left(\sum_{b\in \Theta_1}\beta_{bj}\right)^2}{\sum_{b\in \Theta_1}\beta_{bj}^2},\theta_{1j}=\frac{\sum_{b\in \Theta_1}\beta_{bj}^2}{ \sum_{b\in \Theta_1}\beta_{bj}}.
\end{equation}
To project the interference signal onto the receive beamformer $\bm{w}_{1}$,
(which is a one-dimensional subspace independent of the interfering channel
vector $\bm{h}_{1j}$ of dimension $M B_1$), we again approximate the channel
vector as isotropic.  The projection then results in the scaling of the shape
parameter of the interference as $\frac{k_{1j}}{MB_1}$. Finally, we replace
$B_1$ by its mean $\bar{B}$, then decompose the interference into linear
combination of independent Gamma distributions again using the moment matching
technique as:
\begin{equation}
\nu_{1j}^{(UL)}=|\bm{w}_{1}^H \bm{h}_{1j}|^{2}\approx\sum_{b\in\Theta_1}\beta_{bj}
	G_{bj}^{(\frac{1}{\bar{B}})},
\label{single_IN_UL}
\end{equation}
where $G_{bj}^{(\frac{1}{\bar{B}})}$ are i.i.d. $\Gamma\left(\frac{1}{\bar{B}},1\right)$
distributed. 

The aggregate residual interference is the sum of interference from all
interfering users:
\begin{equation}
\label{inter_dist_UL}
\nu_{1}^{(UL)}=\sum_{j\notin \Omega_1}\nu_{1j} \approx
\sum_{j\notin \Omega_1}\sum_{b\in \Theta_1}\beta_{bj} G_{bj}^{(\frac{1}{\bar{B}})}.
\end{equation}


\begin{figure}
\centering
\includegraphics[width=0.45\textwidth, viewport=70 100 560 340,clip]{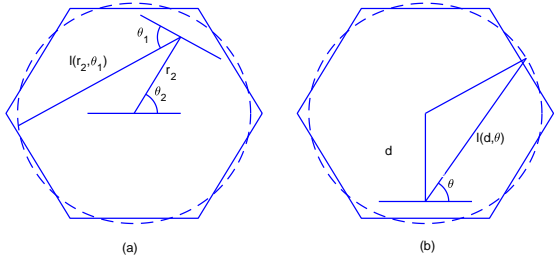}
\caption{Illustration of 2D integration in the evaluation of
    (\ref{laplace_IN_UL}) and
    (\ref{laplace_sig_disjoint_UL_DL})}
\label{2D_integral}
\end{figure}

\subsubsection{Ergodic Rate}

The ergodic rate of the typical user in the UL user-centric network MIMO system
can now be derived using tools from stochastic geometry by using the signal and
interference power distributions (\ref{sig_dist_UL2}) and (\ref{inter_dist_UL})
with further approximations that the cooperation cluster for the typical user is
$\Theta_1 = \Phi_b \cap \mathcal{B}_o\left(R\right)$ and the interfering users
are located in $\Phi_u \setminus \mathcal{B}_o\left(R\right)$.
The achievable rate of the user is computed as $\log(1+ {\rm SINR})$.
By utilizing the following expression of the log function in term of integral
\cite[Lemma 1]{lin2014downlink}
\begin{equation}
\ln(1+x) = \int_0^\infty \frac{e^{-z}}{z} (1- e^{-xz}) \text{d} z,
\end{equation}
the ergodic rate averaged over the distributions of $\Phi_b$ and $\Phi_u$,
can be obtained as follows \cite{lin2014downlink}:
\begin{equation}\label{}
  \bar{C}_U 
	=\int_{0}^{\infty} \frac{e^{-s\sigma^2}}{s} L_{\nu_1^{(UL)}}\left(s\right) \left(1-L_{\zeta_1^{(UL)}}\left(s\right)\right) \text{d}s,
  \label{ergodic_rate_general}
\end{equation}
where $L_{\zeta_1^{(UL)}}\left(s\right)$, $L_{\nu_1^{(UL)}}\left(s\right)$ are respectively
the Laplace transforms of signal and interference power distributions arising from Poisson distributed antenna ports in disjoint regions, as expressed
in (\ref{laplace_sig_UL_DL}) below and in (\ref{laplace_IN_UL}) at the top of the next page:
\begin{equation}
\begin{split}
    &L_{\zeta_1^{(UL)}}\left(s\right)=\mathbb{E}_{\bm{h},\Phi_b}\left[\exp\left(-s\zeta_1^{(UL)}\right)\right]\\
    &=\mathbb{E}_{\bm{h},\Phi_b}\left[\prod_{x_b\in \Phi_b\cap
	\mathcal{B}_o\left(R\right)}\exp\left(-s\beta_{x_b,o} G_{x_b,o}^{(\varpi)}\right)\right]\\
    &\stackrel{\left(a\right)}{=}\mathbb{E}_{\Phi_b}\left[\prod_{x_b\in \Phi_b\cap \mathcal{B}_o
	\left(R\right)}\mathbb{E}_{\bm{h}} \left[\exp\left(-s \beta_{x_b,o}
	G_{x_b,o}^{(\varpi)}\right)\right]\right]\\
    &\stackrel{\left(b\right)}{=}\mathbb{E}_{\Phi_b}\left[\prod_{x_b\in\Phi_b\cap \mathcal{B}_o
	\left(R\right)}\left(1+s\beta_{x_b,o}\right)^{-\varpi}\right]\\
    &\stackrel{\left(c\right)}{=}\exp\left(-\lambda_b\int_{x\in \mathcal{B}_o\left(R\right)} \left(1-\left(1+s\beta_{x,o}\right)^{-\varpi}\right) \text{d}x\right) \\
    &=\exp\left( -2\pi\lambda_b\int_0^R\left( 1-\left(1+s\left(1+\frac{r}{d_0} \right)^{-\alpha}\right)^{-\varpi}\right)\text{d}r\right).\\
    \end{split}
    \label{laplace_sig_UL_DL}
\end{equation}
where (a) comes from the independence of fading and BS location, (b) is based
on the Laplace transform of Gamma random variables, (c) follows from the probability generating functional
(p.g.fl.)  of a PPP \cite{haenggi2012stochastic}.  Here, we use $\beta_{x_b,o}$ to denote
the path-loss component between a BS located at $x_b$ and the typical user
located at the origin, $\beta_{x_b,o}=(1+\frac{|x_b|}{d_0})^{-\alpha}$.
Likewise, $G^{(\varpi)}_{x_b,o}$ denotes the i.i.d. random variable with
$\Gamma(\varpi,1)$ distribution.

\begin{figure}
\centering
\includegraphics[width=0.48\textwidth, viewport=80 100 520 400,clip]{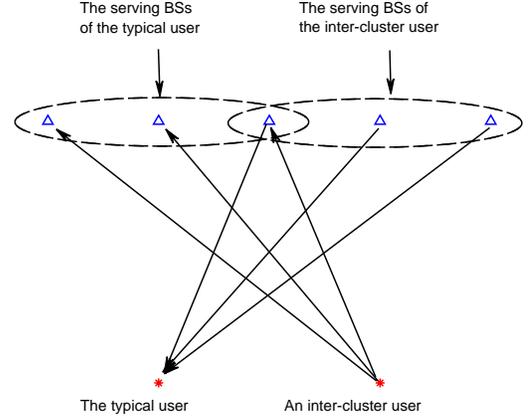}
\caption{Illustration of UL and DL interference in user-centric network MIMO
systems. The distributions of UL and DL interference are the same.}
\label{UL_DL_interference}
\end{figure}


\begin{figure*}[!t]
\normalsize
\begin{equation}
\begin{split}
&L_{\nu_1^{(UL)}}\left(s\right)=\mathbb{E}_{\bm{h},\Phi_b,\Phi_u}\left[\exp\left(-s\nu_1^{(UL)}\right)\right]\\
&\stackrel{\left(a\right)}{=}\mathbb{E}_{\Phi_b}\left[\prod_{x_b\in\Phi_b\cap\mathcal{B}_o\left(R\right)}\mathbb{E}_{\Phi_u}\left[\prod_{x_u\in \Phi_u\setminus
\mathcal{B}_o\left(R\right)}\mathbb{E}_h\left[\exp\left(-s\beta_{x_b,x_u}G^{(\frac{1}{\bar{B}})}_{x_b,x_u}\right)\right]\right]\right]\\
&\stackrel{\left(b\right)}{=}\mathbb{E}_{\Phi_b}\left[\prod_{x_b\in\Phi_b\cap\mathcal{B}_o\left(R\right)}\mathbb{E}_{\Phi_u}\left[\prod_{x_u\in \Phi_u\setminus \mathcal{B}_o\left(R\right)}\left(1+s\left(1+\frac{|x_u-x_b|}{d_0}\right)^{-\alpha}\right)^{-\frac{1}{\bar{B}}}\right]\right]\\
&\stackrel{\left(c\right)}{=}\mathbb{E}_{\Phi_b}\left[\prod_{x_b\in\Phi_b\cap\mathcal{B}_o\left(R\right)}\exp\left(-\lambda_u\int_0^{2\pi}\int_{l\left(|x_b|,\theta_1\right)}^\infty\left(1-\left[1+s\left(1+\frac{r_1}{d_0}\right)^{-\alpha}\right]^{-\frac{1}{\bar{B}}}\right) r_1 \text{d}r_1 \text{d}\theta_1\right)\right]\\
&\stackrel{\left(d\right)}{=}\exp\left(-\lambda_b\int_0^{2\pi}\int_0^{R} \left[1-\exp\left(-\lambda_u\int_0^{2\pi}\int_{l\left(r_2,\theta_1\right)}^\infty\left(1-\left[1+s\left(1+\frac{r_1}{d_0}\right)^{-\alpha}\right]^{-\frac{1}{\bar{B}}}\right) r_1 \text{d}r_1 \text{d}\theta_1 \right)\right] r_2 \text{d}r_2 \text{d}\theta_2\right)\\
&=\exp\left(-2\pi\lambda_b\int_0^{R} \left[1-\exp\left(-\lambda_u\int_0^{2\pi}\int_{l\left(r_2,\theta_1\right)}^\infty\left(1-\left[1+s\left(1+\frac{r_1}{d_0}\right)^{-\alpha}\right]^{-\frac{1}{\bar{B}}}\right) r_1 \text{d}r_1 \text{d}\theta_1 \right)\right]r_2 \text{d}r_2\right)\\
&=\exp\left(-2\pi\lambda_b\int_0^{R}
\left[1-\exp\left(-\lambda_ud_0^2\int_0^{2\pi}P\left(s,\frac{l\left(r_2,\theta_1\right)}{d_0}\right) \text{d}\theta_1 \right)\right]r_2 \text{d}r_2\right),\\
\end{split}
\label{laplace_IN_UL}
\end{equation}
where $l\left(r_2,\theta_1\right)=\sqrt{R^2-r_2^2\cos^2\theta_1}+r_2\sin\theta_1$ and
\begin{multline*}
P\left(s,\frac{l\left(r_2,\theta_1\right)}{d_0}\right)=\frac{-\left(\frac{l\left(r_2,\theta_1\right)}{d_0}+1\right)^2}{2}{}_2F_1\left(-\frac{2}{\alpha},\frac{1}{\bar{B}},1-\frac{2}{\alpha},
-s\left(\frac{l\left(r_2,\theta_1\right)}{d_0}+1\right)^{-\alpha}-1\right)+\\
\left(\frac{l\left(r_2,\theta_1\right)}{d_0}+1\right){}_2F_1\left(-\frac{1}{\alpha},\frac{1}{\bar{B}},1-\frac{1}{\alpha},-s\left(l\left(r_2,\theta_1\right)+1\right)^{-\alpha}-1\right).
\end{multline*}
\hrulefill
\vspace*{4pt}
\end{figure*}


In (\ref{laplace_IN_UL}), the integration is over the 2D plane outside of the
disk $\mathcal{B}_o(R)$ as illustrated in Fig.~\ref{2D_integral}(a). 
The integral involves the hypergeometric function ${}_2F_1\left(\cdot\right)$.
Again, (a) comes from the independence of fading, BS and user location; (b) is
based on the Laplace transform of Gamma random variables; (c) and (d) follow from
the probability generating functional (p.g.fl.) of a PPP \cite{haenggi2012stochastic}.
Here, $\beta_{x_b,x_u}=(1+\frac{|x_b-x_u|}{d_0})^{-\alpha}$;
$G^{(\frac{1}{\bar{B}})}_{x_b,x_u}$ are i.i.d. $\Gamma(\frac{1}{\bar{B}},1)$ distributed.

Note that we have made an implicit assumption that the signal power and the interference power distributions are independent.
Although strictly speaking correlation between signal and interference power distribution exists, since both depend on the BS and user locations, the effect of such an approximation is expected to be minor.

\subsection{Analysis of Downlink}
Examining the SINR expressions (\ref{SINR_UL}) and (\ref{SINR_DL}), it is clear
that the signal component of the typical user has the same form in DL as in UL.
Thus, the signal distribution in DL is the same as that in UL i.e., as in
(\ref{sig_dist_UL2}).
\begin{equation}
    \zeta_1^{(DL)}=\sum_{b\in \Theta_1
}\beta_{b1}G_{b1}^{(\varpi)}.
    \label{sig_random_topo_DL}
\end{equation}



The analysis of the interference is more complicated, but it turns out that
the UL and DL interference distributions have approximately the same expression.
Consider a user $j$ for whom the typical user is its inter-cluster user. 
A careful
observation reveals that the DL interference that is received by the typical user
and is due to the transmission to user $j$, i.e.,
\begin{equation}
    \begin{split}
    \nu_{1j}^{(DL)}=\mid\bm{h}_{j1}^H \bm{w}_{j}\mid^{2}\stackrel{}{\approx}
	\sum_{b\in \Theta_j}\beta_{b1}G_{b1}^{(\frac{1}{\bar{B}})}
    \end{split}
    \label{single_IN_DL}
\end{equation}
has the same form as the corresponding UL interference generated by user $j$
and received at the typical user's cooperating BS cluster, so it has the same
approximate distribution as (\ref{single_IN_UL}) in the uplink case.

The reason for this equivalence of UL and DL interference is illustrated in
Fig.~\ref{UL_DL_interference}.  The UL interference from user $j$ to the
serving BSs of the typical user in the UL is the projection of the UL vector
channel onto the typical user's receive beamformer. The DL interference at the
typical user due to user $j$'s transmission is the projection of the DL vector
channel onto user $j$'s transmit beamformer.  The UL and DL vector channels
have the same statistical distribution because of the symmetry of the channels
and uplink-downlink reciprocity.  The two channels involve the same path-loss
component between a user and a neighboring user's serving BSs as illustrated in
Fig.~\ref{UL_DL_interference}. The UL and DL beamforming vectors also have the
same distribution. Therefore, the overall UL and DL interference distributions
are the same.

The aggregate interference of the typical user in DL is the summation of
interference from all the BSs serving the interfering users:
\begin{equation}
    \begin{split}
    \nu_{1}^{(DL)}=\sum\limits_{j:\; 1 \notin\Omega_j}\nu_{1j}^{(DL)}=
\sum\limits_{j:\; 1 \notin\Omega_j}
\sum_{b\in\Theta_j}\beta_{b1}G_{b1}^{(\frac{1}{\bar{B}})}.
    \end{split}
\end{equation}
When the user and BS locations are distributed as PPPs, and if we assume that
the interfering users are simply ones outside of the circle with radius $R$,
the aggregate DL interference at the typical user due to the transmission for all
DL interfering users must also have the same distribution as the aggregate UL
interference received by the typical user's serving BSs due to the transmissions
by all UL interfering users.

Finally, as both the signal and interference in the UL and DL have the same
distributions, so must their Laplace transforms.
Consequently, the ergodic rate of the typical user in DL must have the same
expression as the ergodic rate of the typical user in UL, as obtained by
plugging (\ref{laplace_sig_UL_DL}) and (\ref{laplace_IN_UL}) into
(\ref{ergodic_rate_general}). The only difference is that the normalized
background noise variance $\sigma_u^2$ in UL should be replaced by $\sigma_d^2$
in DL, accounting for the difference in transmit power budgets in the UL and DL.

\section{Stochastic Analysis of Disjoint Clustering}


One of the goals of this paper is to compare user-centric clustering with
baseline disjoint
clustering. Toward this end, we derive in this section a stochastic analysis of
disjoint clustering. In our model of baseline disjoint clustering, the network is
partitioned into hexagonal regions; BSs within the hexagon form cooperative
clusters. The hexagon is further approximated as a circle with
the same area to facilitate analysis. To make comparison with user-centric
clustering, we set the cluster radius $R$ to be the same in both cases.


We follow the same methodology for analysis as in the previous section. In
fact, the analysis of DL disjoint clustering has already been carried out in
\cite{Kianoush_Cooperation_gain}; the main contribution of
this section is the corresponding UL analysis. For both UL and DL, the main
difference between disjoint and user-centric clustering is that the serving BSs
are no longer symmetrically centered around the user.  Thus, the analysis needs to
be modified in two respects.  First, the computation of achievable rate for a
user depends on its location; it requires a careful analysis of the
distribution of the channel strengths from the serving BSs. Second, because the rates are
location dependent, average rate is no longer necessarily an adequate measure.
Cluster-edge performance is of equal, if not more, importance.

We use $\bar{C}_{D}\left(d\right)$ to denote the ergodic rate of a user whose
distance to the cluster center is $d$.
The average user rate under disjoint clustering can be derived as
\begin{equation}
\bar{C}_D=\int_0^R f\left(d\right) \bar{C}_{D} \left(d\right) \;\rm{d} \emph{d},
\label{disjoint_perfor_metric}
\end{equation}
where $f\left(d\right)$ is the probability density function of the distance $d$
of the given user. Assume that users are uniformly distributed within the
circle with radius $R$, then 
$f\left(d\right)=\frac{2d}{R^2}$. It remains to find $\bar{C}_D \left(d\right)$ in both UL and DL.

\subsection{Ergodic Rate of Location-Specific User in the UL}

Without loss of generality, we consider the ergodic rate of a location-specific
user, user $i$, in the cluster centered at the origin, located at $x_0$ with distance $|x_0|=d$ from the
origin. In the UL, the signal power depends on the distance between the user
and its serving BSs. Using the same idea as in the user-centric case, the
signal power can be approximated as a summation of Gamma random variables
over the locations of its serving BSs as follows:
\begin{equation}
    \zeta_{i}^{(UL)}=\sum_{x_b\in \Phi_b\cap
	\mathcal{B}_o\left(R\right)}\beta_{x_b,x_0} G_{x_b,x_0}^{(\varpi)}.
\label{sig_random_disjoint_UL}
\end{equation}
The derivation of the above result is similar to the analysis of signal power
in the user-centric case in (\ref{sig_dist_UL2}); the key difference is that
user-location specific distances need to be accounted for.

As far as interference is concerned, it is not difficult to see that the UL
interference for the user-centric and the disjoint clustering cases have
identical distributions, i.e., as expressed in (\ref{inter_dist_UL}).

Thus, the ergodic rate of a location-specific user can be derived as follows
\cite{lin2014downlink},
\begin{equation}
  \bar{C}_D \left(d\right)=\int_{0}^{\infty} \frac{e^{-s\sigma^2}}{s}
L_{\nu_{i}^{(UL)}}\left(s\right) \left(1-L_{\zeta_{i}^{(UL)}}\left(s\right)\right) \text{d}s,
  \label{ergodic_rate_formula_UL_disjoint}
\end{equation}
where $L_{\nu_{i}^{(UL)}}\left(s\right) = L_{\nu_{1}^{(UL)}}\left(s\right)$ is the Laplace
transform of the interference distribution as already computed in
(\ref{laplace_IN_UL}), and $L_{\zeta_{i}^{(UL)}}\left(s\right)$ is the Laplace
transform of the signal distribution as:
\begin{equation}
\begin{split}
    &L_{\zeta_{i}^{(UL)}}\left(s\right)=\mathbb{E}_{\bm{h},\Phi_b}\left[\exp\left(-s\zeta_{i}^{(UL)}\right)\right]\\
    &=\mathbb{E}_{\bm{h},\Phi_b}\left[\prod_{x_b\in \Phi_b\cap
\mathcal{B}_o\left(R\right)}\exp\left(-s\beta_{x_b,x_0}G_{x_b,x_o}^{(\varpi)}\right)\right]\\
    &\stackrel{(a)}{=}\mathbb{E}_{\Phi_b}\left[\prod_{x_b\in \Phi_b\cap
\mathcal{B}_o\left(R\right)}\mathbb{E}_{\bm{h}}
\left[\exp\left(-s\beta_{x_b,x_0}G_{x_b,x_0}^{(\varpi)}\right)\right]\right]\\
    &\stackrel{(b)}{=}\mathbb{E}_{\Phi_b}\left[\prod_{x_b\in\Phi_b\cap \mathcal{B}_o\left(R\right)}\left(1+s\beta_{x_b,x_0}\right)^{-\varpi}\right]\\
    &\stackrel{(c)}{=}\exp\left(-\lambda_b\int_{x\in\mathcal{B}_o\left(R\right)} \left(1-\left(1+s\beta_{x,x_0}\right)^{-\varpi}\right) \text{d}x\right) \\
    &=\exp\left(-\lambda_b\int_0^{2\pi}\int_0^{l\left(d,\theta\right)} \right.  \\
& \qquad \qquad \left. 
\left( 1 -  \left(1+s\left(1+\frac{r}{d_0} \right)^{-\alpha}\right)^{-\varpi}\right)r \text{d}r\text{d}\theta\right),\\
    \end{split}
    \label{laplace_sig_disjoint_UL_DL}
\end{equation}
where again $l\left(d,\theta\right)=\sqrt{R^2-d^2\cos^2\theta}+d\sin\theta$ is
used in the integration to account for the location-specific user, as
illustrated in Fig.~\ref{2D_integral}(b).

\begin{table*}[t]
  \centering
  \caption{The signal and interference distributions in the UL and DL under user-centric and disjoint clustering}
  \begin{tabular}{|>{\centering\arraybackslash}m{0.6in}|>{\centering\arraybackslash}m{0.5in}|>{\centering\arraybackslash}m{1.5in}|>{\centering\arraybackslash}m{2.2in}|}
\hline
& \rule{0pt}{3ex} & Signal Distribution & Interference Distribution\\
\hline
\multirow{2}{*}{User-centric} \rule{0pt}{6ex} & Uplink & $\sum\limits_{x_b\in
\Phi_b\cap
\mathcal{B}_o\left(R\right)}\beta_{x_b,o}G_{x_b,o}^{\left(\varpi\right)}$ &
$\sum\limits_{x_u\in\Phi_u\setminus
\mathcal{B}_o\left(R\right)}\sum\limits_{x_b\in \Phi_b\cap
\mathcal{B}_{o}\left(R\right)}\beta_{x_b,x_u}G_{x_b,x_u}^{(\frac{1}{\bar{B}})}$\\
\cline{2-4}
\rule{0pt}{6ex}
& Downlink & $\sum\limits_{x_b\in \Phi_b\cap \mathcal{B}_o\left(R\right)}\beta_{x_b,o}
	G_{x_b,o}^{\left(\varpi\right)}$ & $\sum\limits_{x_u\in\Phi_u\setminus \mathcal{B}_o\left(R\right)}\sum\limits_{x_b\in \Phi_b\cap \mathcal{B}_{x_u}\left(R\right)}\beta_{x_b,o}
	G_{x_b,x_u}^{(\frac{1}{\bar{B}})}$\\
\hline
\multirow{2}{*}{Disjoint}
\rule{0pt}{6ex}
& Uplink & $\sum\limits_{x_b\in \Phi_b\cap \mathcal{B}_o\left(R\right)}\beta_{x_b,x_0}
	G_{x_b,o}^{\left(\varpi\right)}$ & $\sum\limits_{x_u\in \Phi_u\setminus \mathcal{B}_o\left(R\right)}\sum\limits_{x_b\in \Phi_b\cap \mathcal{B}_o\left(R\right)}\beta_{x_b,x_u}
	G_{x_b,x_u}^{(\frac{1}{\bar{B}})}$\\
\cline{2-4}
\rule{0pt}{6ex}
& Downlink & $\sum\limits_{x_b\in \Phi_b\cap \mathcal{B}_o\left(R\right)}\beta_{x_b,x_0}
	G_{x_b,o}^{\left(\varpi\right)}$ & $\sum\limits_{x_b\in \Phi_b\setminus \mathcal{B}_o\left(R\right)}\beta_{x_b,x_0}
	G_{x_b}^{(K)}$\\
\hline
\end{tabular}
\label{summary_table}
\end{table*}

\subsection{Ergodic Rate of the Location-Specific User in the DL}

The ergodic rate of the location-specific user in the DL can be derived in a similar way.
The rate expression has already been derived in \cite{Kianoush_Cooperation_gain}.
We briefly summarize the result below.

First, the signal power distribution in the DL and UL are identical, i.e., as
(\ref{sig_random_disjoint_UL}) with a Laplace transform as expressed in
(\ref{laplace_sig_disjoint_UL_DL}).
Second, the interference power can be written in the following form.
\begin{equation}
\nu_{i}^{(DL)}=\sum_{x_b\in \Phi_b\setminus
{\mathcal{B}_o}\left(R\right)}\beta_{x_b,x_0}G_{x_b,x_0}^{(K)}.
\label{inter_dist_disjoint_DL}
\end{equation}
where $G_{x_b,x_0}^{(K)}$ are i.i.d. distributed as $\Gamma(K,1)$.

We see that the interference is the summation of the path-loss component from
all the out-of-cluster BSs to the location-specific user, multiplied by a Gamma
random variable of shape parameter $K$. The difference between this interference
expression and the interference distribution in the other three cases, i.e.,
(\ref{inter_dist_UL}), reflects the different interfering paths in this disjoint
DL case.  A detailed derivation of this result is available in
\cite{Kianoush_Cooperation_gain}.
The Laplace transform of the distribution in (\ref{inter_dist_disjoint_DL}) can
be expressed as follows:
\begin{equation}
\begin{split}
&L_{\nu_{i}^{(DL)}}\left(s\right)=\mathbb{E}_{\bm{h},\Phi_b}\left[\exp\left(-s \nu_{i}^{(DL)}\right)\right]\\
&=\exp\left(-\lambda\int_0^{2\pi}\int_{l\left(d,\theta\right)}^\infty \right.  \\
& \qquad \qquad \left.
\left(1-\left(1+s\left(1+\frac{r}{d_0}\right)^{-\alpha}\right)^{-K}\right) r \text{d}r \text{d}\theta\right).\\
\end{split}
\label{laplace_IN_UL_DL}
\end{equation}
Finally, the ergodic rate of the location-specific user is obtained by
substituting $L_{\zeta_{i}^{(UL)}}\left(s\right)$ and $L_{\nu_{i}^{(UL)}}\left(s\right)$ in
(\ref{ergodic_rate_formula_UL_disjoint}) with (\ref{laplace_sig_disjoint_UL_DL})
and (\ref{laplace_IN_UL_DL}).



\section{Analytic Comparison of User-Centric and Baseline Disjoint Clustering}

The analytic results on the signal and interference distribution derived in the
previous section are summarized in Table \ref{summary_table}. These analytic
results illustrate the advantage of user-centric clustering as compared to disjoint
clustering.

For the received signal power distribution, we see that in both the
user-centric and disjoint clustering, the received signal power is a linear
combination of Gamma random distributions of the same parameter, but the
coefficients of linear combination differ. This is due to the fact that the
user is located at the center of the cluster in the user-centric case, so its
serving BSs are located symmetrically around the user, while in the disjoint case the
analysis of path-losses from the serving BSs is a function of the user location's 
distance $d$ from the cluster
center. Thus, user-centric clustering has benefit in term of signal power. This
applies to both UL and DL, (in fact UL and DL powers have the same distribution).

For the interference power distribution, first we see that the UL user-centric
and disjoint cases have the same distribution. For the DL, although the DL
interference in the user-centric case has a different form as in UL, its
distribution actually turns out to be identical. The only case where the
interference distribution is different is the DL disjoint case. Here, the Gamma
random variable has a different shape parameter.
Interestingly, the coefficients of
linear combination in the DL disjoint case involves only BSs outside of the
cluster, while in the DL user-centric case it involves BSs inside the cluster
as well. 
In fact, the DL interference is improved in the user-centric case as compared
to the disjoint case for cluster-edge users (where $d$ is large), but is
worsened for cluster-center users (where $d$ is small).

As consequence, user-centric clustering is expected to outperform disjoint
clustering uniformly in the UL, but in the DL, the main advantage of
user-centric clustering is for cluster-edge users.

%

\section{Numerical Comparison of User-Centric and Baseline Disjoint Clustering}

\begin{figure}
\centering
\includegraphics[width=0.5\textwidth, trim=30 20 20 10, clip]{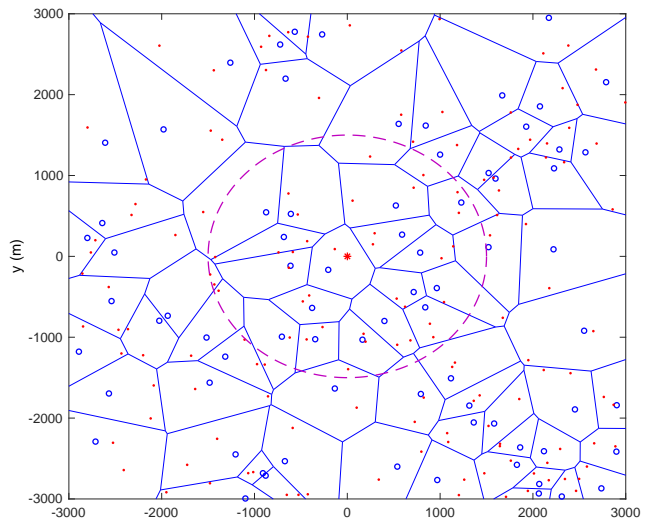}
\caption{A stochastic model of cellular networks where the BSs and the users are
distributed as a Poisson process.}
\label{Poisson_cellular_model}
\end{figure}

This section provides numerical results to validate the analysis developed in
this paper.  We consider a stochastic deployment of BSs in a cellular network
as shown in Fig.~\ref{Poisson_cellular_model}. Each BS is equipped with
4 antennas and schedules 2 single-antenna users (in a round-robin fashion) within
its voronoi cell, i.e., $M=4, K=2$, with loading factor 0.5.  The power of the
transmit beam for each user in the DL is set to be 40 dBm over 20MHz bandwidth.
The transmit power of each user in UL is 23dBm over 20MHz.  Power spectrum
density of the background noise is set to -174dBm/Hz; a noise figure of 9dB and an SINR gap of 3dB are
included. We use a path-loss model of $128.1+37.6 \log (d)$ in dB, where $d$ is
expressed in km.  This corresponds to a path-loss exponent of 3.76 with
reference distance $d_0$ approximately equal to 0.3920m.

\subsection{Signal and Interference Power Distributions}

\begin{figure}
\centering
\includegraphics[width=0.5\textwidth,trim=40 10 20 20, clip]{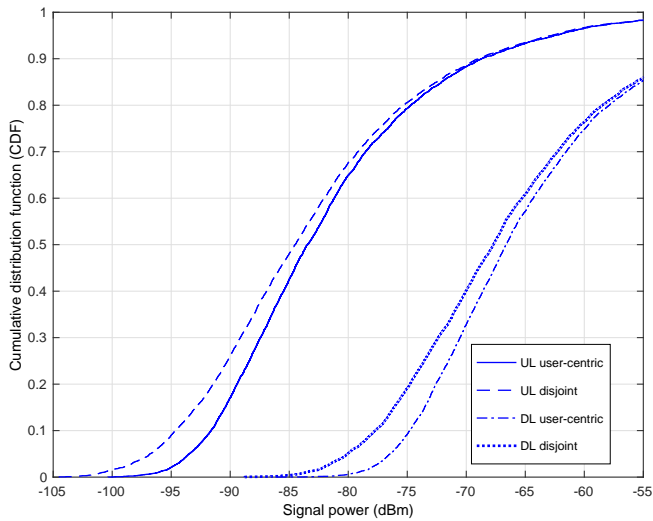}
\caption{Comparison of UL and DL signal power: User-centric vs. disjoint clustering}
\label{signal_comparison}
\end{figure}

We first validate the qualitative comparison of signal and interference power
by plotting their distributions for both UL and DL for a fixed scenario with
average cluster size of 6 BSs. Fig.~\ref{signal_comparison} shows the signal
power distributions of the user-centric vs. disjoint clustering from simulation. The
signal power in DL is larger than UL due to the larger DL transmission power.
The benefit of user-centric clustering is clearly shown, especially for the
cluster edge users.

\begin{figure}
\centering
\includegraphics[width=0.5\textwidth,trim=40 10 20 20, clip]{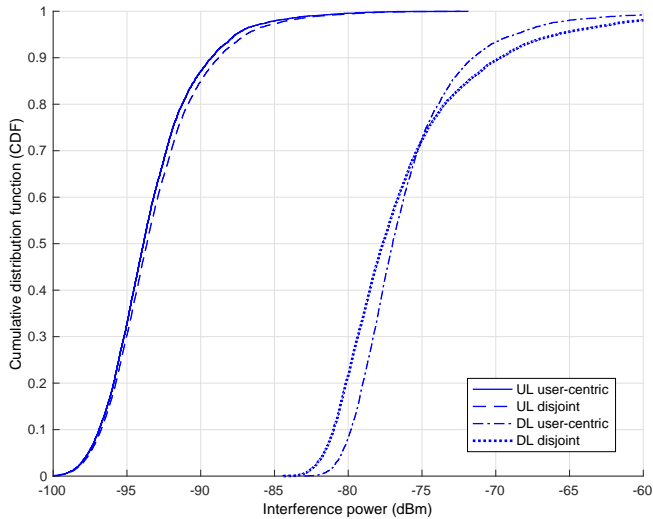}
\caption{Comparison of UL and DL interference power: User-centric vs. disjoint clustering}
\label{interference_comparison}
\end{figure}

Fig.~\ref{interference_comparison} shows that interference power distributions
in user-centric clustering as compared to disjoint clustering. We see that the
two cases are nearly identical in the UL. In the DL, the interference power in the
user-centric case is smaller than the disjoint case for cluster-edge users (i.e.,
users with strong interference), but stronger for cluster-center users (i.e.,
users with weak interference). This agrees with the analytic comparison.
User-centric clustering improves the cluster-edge user performance at slight
expense of cluster-center users.

\subsection{Ergodic Rate}

\begin{figure}
\centering
\includegraphics[width=0.5\textwidth,trim=40 10 20 20, clip]{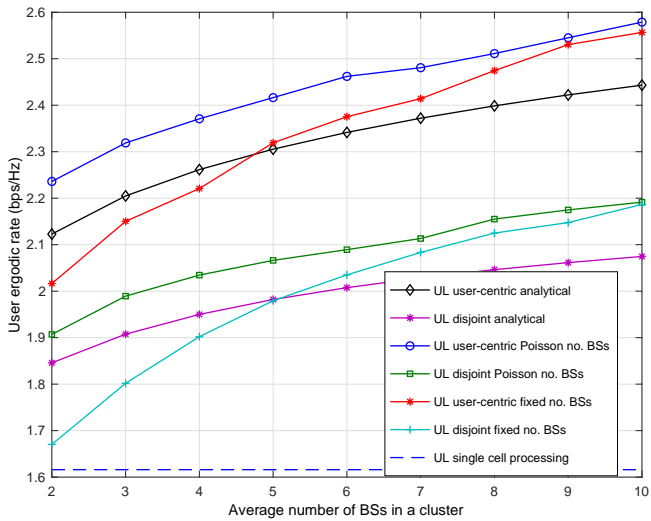}
\caption{Ergodic user rate for UL with disjoint and user-centric clustering as
function of cooperation cluster size. The cases with both fixed number or Poisson
distributed number of serving BSs are plotted. Single-cell processing case is
included as a reference.}
\label{simu_ergodic_UL}
\end{figure}

\begin{figure}
\centering
\includegraphics[width=0.5\textwidth,trim=40 10 20 20, clip]{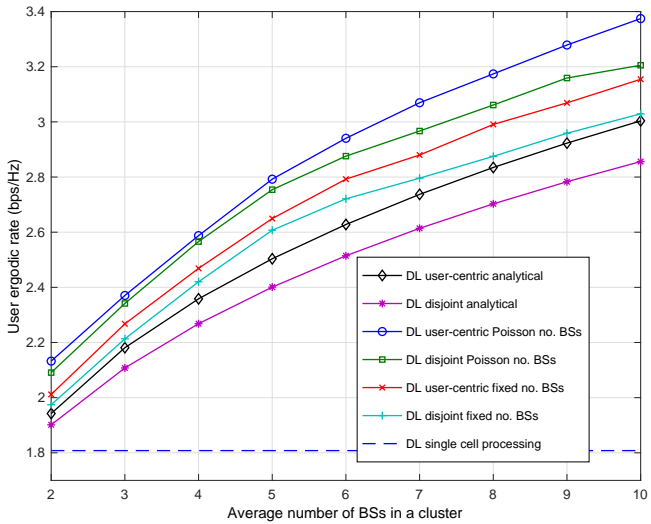}
\caption{Ergodic user rate for DL with disjoint and user-centric clustering as
function of cooperation cluster size. The cases with both fixed number or Poisson
distributed number of serving BSs are plotted. Single-cell processing case is
included as a reference.}
\label{simu_ergodic_DL}
\end{figure}

Next we investigate the ergodic rate of the typical user in both UL and DL as
a function of cluster size. Figs.~\ref{simu_ergodic_UL} and \ref{simu_ergodic_DL}
show respectively the UL and DL ergodic user rates evaluated from the analytical
expressions as well as obtained from system-level simulation for both
user-centric and disjoint clustering cases.  The horizontal axis is the average
number of serving BSs in the typical user's cluster, i.e., $\bar{B}$, which is
a measure of cluster size. Recall that the analytic expressions developed in
this paper are based on a fixed cooperation cluster radius $R$.
As BSs are modeled as PPP, the number of BSs within the cluster is a Poisson
random variable. Practical implementation of network MIMO system would likely
have fixed number of BSs in the cooperation cluster. For numerical comparison,
we thus include both the simulation results with Poisson number of BSs as well
as the case with fixed number of exactly $\bar{B}$ BSs in the cooperation cluster.
Finally, the single-cell processing baseline case is also included, where BSs
do not cooperate and each user is served by its nearest BS.

First, we observe that the analytical results for the ergodic sum rate match with the simulation
within an error of about 5\%, which is remarkable given the number of
approximations involved in modeling both the signal and interference. We also
see that the Poisson assumption gives reasonable ergodic rate estimates as
compared to the case where the number of BSs in the cluster is fixed. The cases
with fixed number of BSs vs. Poisson number of BSs are expected to converge
when $\bar{B}$ is large. The simulations show that even at small $\bar{B}$,
the two are fairly close.


We make the following observations on the ergodic rate performance of
user-centric network MIMO systems:
\begin{itemize}
\item The ergodic rate increases as the cluster size grows for both UL and DL
and for both user-centric and disjoint clustering cases.  But larger cluster
benefits in DL more than UL. This is because DL transmit power is larger, so
it is more interference limited than UL. Consequently, interference mitigation
brings more improvement to the DL.
\item User-centric clustering achieves higher ergodic rate than disjoint
clustering for both UL and DL. The benefit of user-centric cluster in UL is
about 15-20\%, and in DL only about 5\%. The main advantage of user-centric
clustering is the enhanced signal power, because user-centric clustering puts
every user at the center of the BS cooperation cluster. Also, as compared to
disjoint clustering, user-centric cooperation reduces interference uniformly
for the UL. But in DL, user-centric clustering reduces interference for
cluster-edge users only, and may actually increase interference for
cluster-center users. Thus, in term of \emph{average rate}, the benefit
of user-centric clustering mainly occurs in the UL.
\end{itemize}





\subsection{Distribution of Rates}

\begin{figure}
\centering
\includegraphics[width=0.5\textwidth,trim=40 10 20 20, clip]{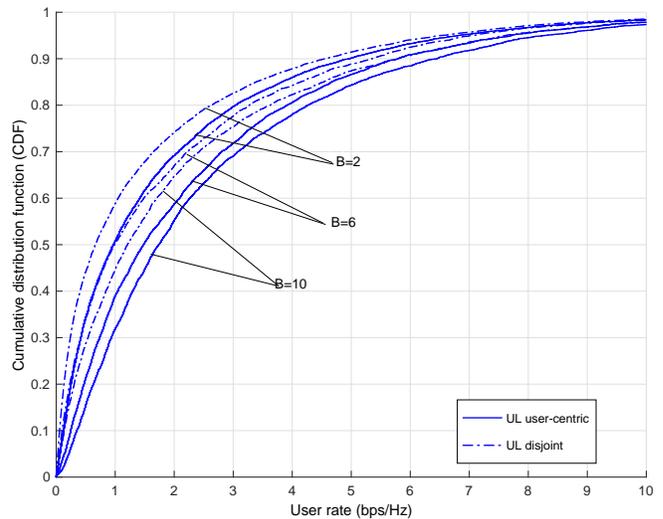}
\caption{User rate distributions for user-centric/disjoint clustering with different cluster sizes in UL}
\label{CDF_UL}
\end{figure}

\begin{figure}
\centering
\includegraphics[width=0.5\textwidth,trim=40 10 20 20, clip]{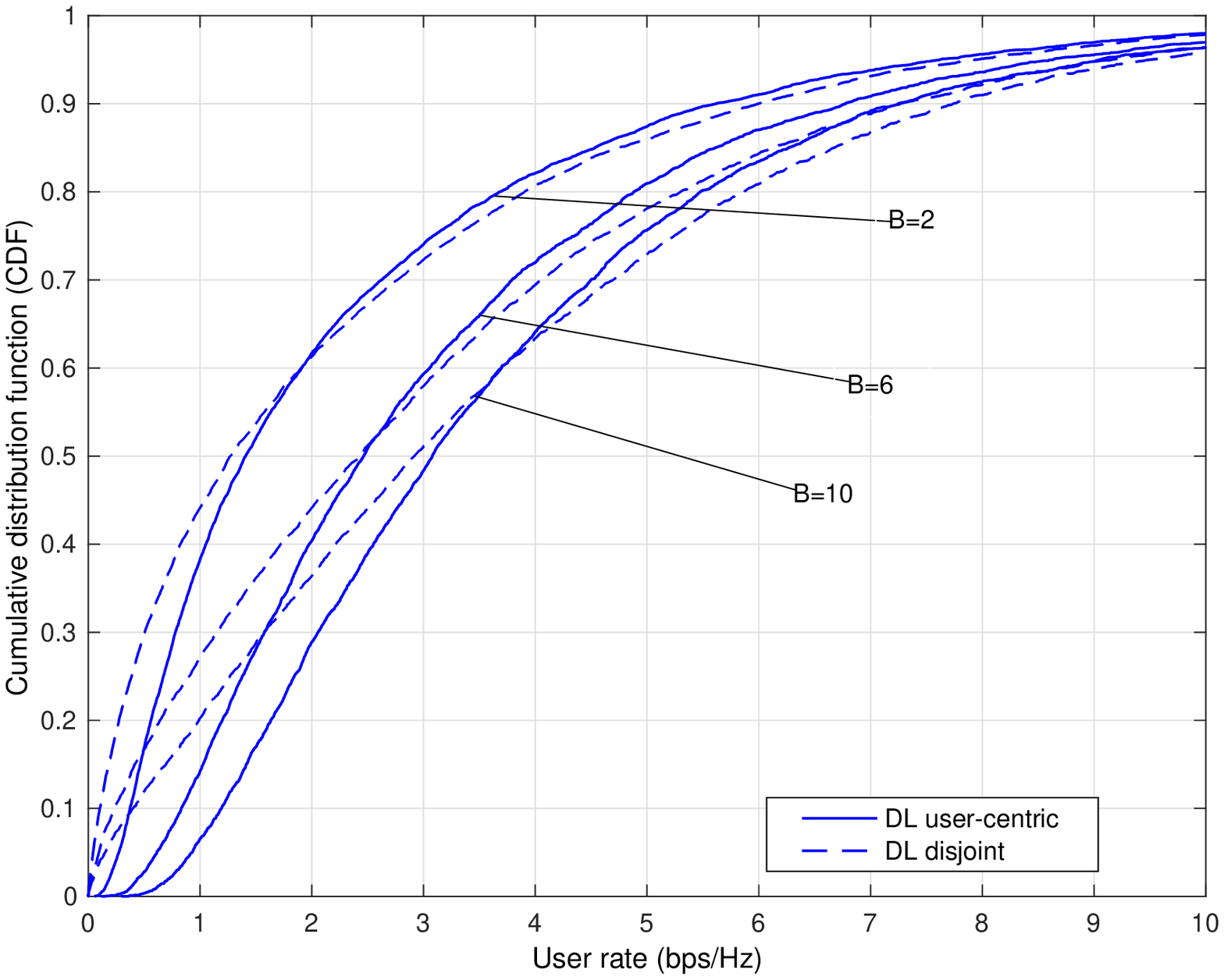}
\caption{User rate distributions for user-centric/disjoint clustering with different cluster sizes in DL}
\label{CDF_DL}
\end{figure}

Although user-centric clustering brings in more average rate improvement in the
UL, it does provide significant improvement to cluster-edge performance in DL.
This is evident from the simulation results of the cumulative distribution function (CDF) shown in
Figs.~\ref{CDF_UL} and \ref{CDF_DL}.

In the CDF curve for the UL case shown in Fig.~\ref{CDF_UL}, we see that the
user-centric CDF curves are always to the right of the disjoint CDF curves.
This is consistent with our analysis: user-centric clustering improves the
performance of every user regardless of the location.

In the CDF curve for the DL case shown in Fig.~\ref{CDF_DL}, we see that the
performance of low-rate users (corresponding to the cluster-edge users in the
disjoint clustering case) is significantly improved, while the performance of
high-rate users is actually reduced under user-centric clustering. This also
confirms our analysis: user-centric clustering improves signal power and
reduces inter-cluster interference for cluster-edge users, while creating
additional intra-cluster interference for cluster-center users.

It should be noted that the improvement for cluster-edge users for DL user-centric
clustering is significant. The 10-percentile user rate performance is improved
by a factor of three or more at $\bar{B}=10$.  Given the importance of low-rate
user performance to the cellular service providers, this provides strong
justification for the user-centric architecture.



\begin{figure}
\centering
\includegraphics[width=0.5\textwidth,trim=40 10 20 20, clip]{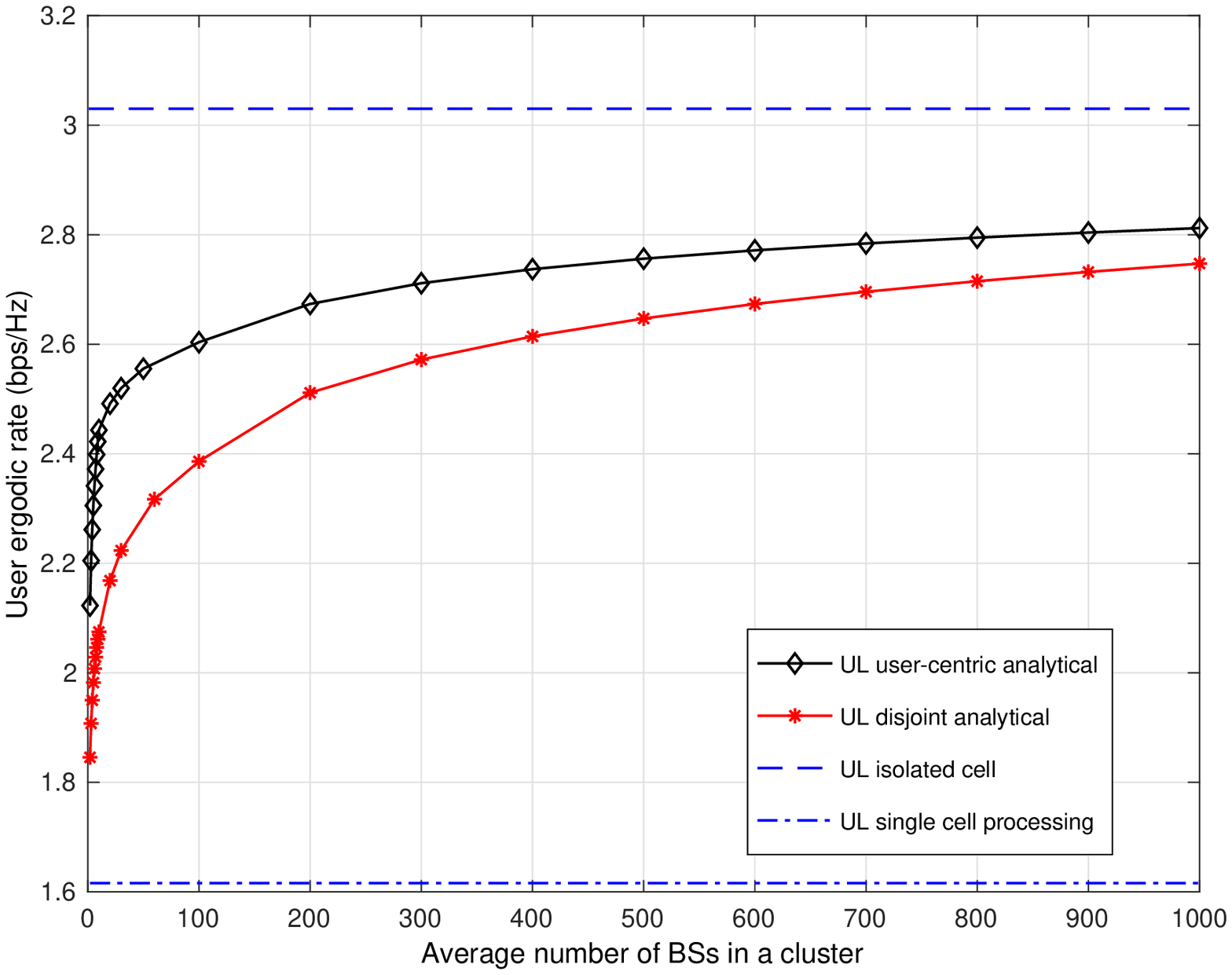}
\caption{Ergodic rate under user-centric/disjoint clustering with different
	cluster sizes in UL, with isolated-cell case included as reference}
\label{large_cluster_UL}
\end{figure}

\begin{figure}
\centering
\includegraphics[width=0.5\textwidth,trim=40 10 20 20, clip]{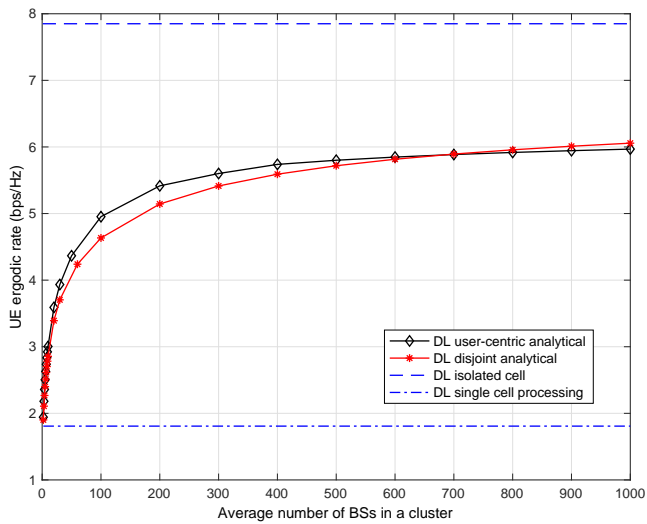}
\caption{Ergodic rate under user-centric/disjoint clustering with different
	cluster sizes in DL, with isolated-cell case included as reference}
\label{large_cluster_DL}
\end{figure}

\subsection{Asymptotic Performance}

To obtain a better understanding of the role of BS cooperation,
Figs.~\ref{large_cluster_UL} and \ref{large_cluster_DL} plot the
analytic ergodic rate expression for user-centric and disjoint clustering
in the limit as the number of BSs in the cluster goes to infinity. (For large
cluster size, simulation would not have been computationally feasible to do.)
The results show:

\begin{itemize}

\item BS cooperation brings more improvement in the DL as compared to UL as
compared to single-cell processing.
Cluster size of $\bar{B}=10$ achieves 20\% gain in UL and 40\% in DL.
Cluster size of $\bar{B}=100$ achieves 60\% gain in UL and 150\% in DL.
The difference is due to the fact that DL is more interference limited than UL.

\item User-centric clustering always outperforms disjoint clustering in the UL,
but the gap between the two narrows at large cluster size. In the DL,
user-centric clustering does not have a significant benefit in term of ergodic
rate as compared to disjoint clustering; (the main advantage of DL user-centric
clustering is for cluster-edge performance.)

\item Even as cluster size goes to infinity (which is clearly not
realistic, given the amount of backhaul signalling that would require),
the analytic performance of network MIMO system with either user-centric or disjoint
clustering derived in this paper does not approach that of an isolated cell. Thus, there is
significant signalling cost to BS cooperation (at least with ZFBF), even
before additional overhead, e.g., CSI acquisition cost, is taken into account.
\end{itemize}


\section{Conclusion}

This paper analyzes the system performance of network MIMO system with
zero-forcing beamforming across multiple BSs, with either user-centric and
disjoint BS clustering, for both DL and UL. Using tools from stochastic geometry
and by approximating the channel and interference power distributions, we derive
tractable analytical expressions of ergodic rate and reveal the benefit of
cooperation in different cases.  This paper shows that as compared to the
baseline na\"{i}ve disjoint clustering strategy, user-centric clustering
brings advantage to cluster-edge users in DL network MIMO systems, while
improving user rates uniformly in the UL. The analytic techniques developed in
this paper provide useful insight to the design of future cooperative wireless
cellular communication networks.



\nocite{zhu_spawc}

\bibliographystyle{IEEEtran}

\bibliography{IEEEabrv,user_centric}

\end{document}